\documentstyle[11pt,moriond]{article}
\newcount\Mac  \Mac=0  
\newcommand{\ifMac}[2]{\ifnum\Mac=1#1\else#2 \fi}
\newcount\compatto  \compatto=1  
\newcommand{\cQ}[2]{\ifnum\compatto=1#1\else#2\fi}

\def\putps(#1,#2)(#3,#4)#5#6{\ifnum\Mac=1 \put(#1,#2){\special{picture #5}}
\else  \put(#3,#4){\includegraphics{#6}} \fi}
\def\Red  {}

\def\Black{}
\def\Orange{} 
\def\Green{} 
\def\GreenDark{} 
\def\Blue {}

\newcommand{\eps}{\varepsilon}

\newcommand{\GeV}{\,{\rm GeV}}
\newcommand{\TeV}{\,{\rm TeV}}

\newcommand{\MGUT}{M_{\rm GUT}}
\newcommand{\MM}{M_{\rm GM}}

\newcommand{\One}{\hbox{1\kern-.24em I}}
\newcommand{\NP}{Nucl. Phys.}

\newcommand{\PL}{Phys. Lett.}
\newcommand{\PR}{Phys. Rev.}

\newcommand{\SUSY}{\hbox{``SUSY''}}
\newcommand{\LEP}{\hbox{``LEP''}}

\def\Ord{{\cal O}}  \def\SU{{\rm SU}}
 
\def\circa#1{\,\raise.3ex\hbox{$#1$\kern-.75em\lower1ex\hbox{$\sim$}}\,}
\makeatletter
\def\art{\@ifnextchar[{\eart}{\oart}}
\def\eart[#1]#2#3#4#5#6{{\rm #2}, {\em #3 \bf #4} {\rm (#6) #5} ({\em #1})}
\def\hepart[#1]#2{{\rm #2, \em#1}}
\newcommand{\oart}[5]{{\rm #1}, {\em #2 \bf #3} {\rm (#5) #4}}
\newcommand{\y}{{\rm and} }
\newcounter{alphaequation}[equation]
\def\thealphaequation{\theequation\hbox to
0.6em{\hfil\alph{alphaequation}\hfil}}
\def\eqnsystem#1{
\def\@eqnnum{{\rm (\thealphaequation)}}
\def\@@eqncr{\let\@tempa\relax \ifcase\@eqcnt \def\@tempa{& & &} \or
  \def\@tempa{& &}\or \def\@tempa{&}\fi\@tempa
  \if@eqnsw\@eqnnum\refstepcounter{alphaequation}\fi
\global\@eqnswtrue\global\@eqcnt=0\cr}
\refstepcounter{equation} \let\@currentlabel\theequation \def\@tempb{#1}
\ifx\@tempb\empty\else\label{#1}\fi
\refstepcounter{alphaequation}
\let\@currentlabel\thealphaequation
\global\@eqnswtrue\global\@eqcnt=0 \tabskip\@centering\let\\=\@eqncr
$$\halign to \displaywidth\bgroup \@eqnsel\hskip\@centering
$\displaystyle\tabskip\z@{##}$&\global\@eqcnt\@ne
\hskip2\arraycolsep\hfil${##}$\hfil& \global\@eqcnt\tw@\hskip2\arraycolsep
$\displaystyle\tabskip\z@{##}$\hfil
\tabskip\@centering&\llap{##}\tabskip\z@\cr}
\def\endeqnsystem{\@@eqncr\egroup$$\global\@ignoretrue} \makeatother

\newcommand{\pausa}{$$
\bullet~
\bullet~
\bullet~
\bullet~
\bullet~
\bullet~
\bullet~
\bullet~
\bullet~
\bullet~
\bullet~
\bullet~
\bullet~
\bullet~
\bullet~
\bullet~
\bullet~
\bullet\Black
$$}

\newcount\n

\begin{document}

\cQ{\centerline{5 Apr.\ 1999 \hfill hep-ph/9904247}}{}
\vspace*{4cm}

\title{\Red\cQ{\LARGE\bf Naturalness of supersymmetric models\vspace{1cm}}{NATURALNESS OF SUPERSYMMETRIC MODELS}}

\author{{\sc\Black\cQ{\large\bf}{} Alessandro Strumia} \cQ{\vspace{5mm}}{}}

\address{\cQ{\large\em}{} Dipartimento di Fisica, Universit\`a di Pisa\\
ed INFN, sezione di Pisa,  I-56126 Pisa, Italia\cQ{\vspace{1cm}}{}}

\maketitle\abstracts{\Blue\cQ{\large}{}
After presenting a simple procedure for testing naturalness
(similar to Bayesian inference and not more subjective than it)
we show that LEP2 experiments pose a naturalness
problem for `conventional' supersymmetric models.
About $95\%$ of the parameter space of minimal supergravity MSSM is excluded by LEP2 experiments.
Moreover in this model electroweak baryogenesis,
or detectable supersymmetric corrections to mixing of $K$ and $B$ mesons,
are possible only in very small corners of the parameter space.
The naturalness problem is stronger in gauge mediation models, expecially with light messengers.
\cQ{\\\phantom{.}~~~}{}We recall some possible explanations (different from an improbable numerical accident)
of why supersymmetry has not (yet?) been found.}\Black

\cQ{\vfill\pausa
\begin{centering}
From a talk presented at the XXXIV Recontres de Moriond on\\

`Electroweak interactions and unified theories', Les Arcs, (13--20)/3/1999.\\

Transparencies avaible at the www address {\tt moriond.in2p3.fr/EW/transparencies.}\\
\end{centering}
\newpage}{}

\section{Introduction}
Naturalness is the only problem of the SM that requires new physics at energies $\Lambda$
accessible to accelerators, $\Lambda\circa{<} 4\pi M_Z$.
Supersymmetry (SUSY) is one possible solution to the problem:
in this case the scale of new physics is given by the mass of the supersymmetric particles
(`sparticles'), that are expected to be just around the $Z$-boson:
$m_{\rm SUSY}=(\frac{1}{3}\div 3)M_Z$.

However experiments now say that most of the sparticles must be heavier than the $Z$ boson.
It is thus interesting to studying if the existing `conventional' supersymmetric models 
(supergravity and gauge mediation)
prefer a chargino mass $M_\chi$ around $\sim 30\GeV$ or $\sim 300\GeV$.
We would like to discuss the following questions:
\cQ{
\begin{quotation}

Experiments are saying that naturalness is a problem for supersymmetry?

If yes, can we found a solution and learn something about the sparticle spectrum?

\end{quotation}}
{$\bullet$ Experiments are saying that naturalness is a problem for supersymmetry?
$\bullet$ If yes, can we found a solution and learn something about the sparticle spectrum?}

\section{Naturalness and fine-tuning}
Usually questions of this kind are discussed using fine tuning (`FT')\cite{FT}
as a quantitative measure of naturalness.
In a supersymmetric model the $Z$ boson mass squared, $M_Z^2(\wp)$, can be computed
as function of the parameters $\wp$ ($\mu$-term, gaugino masses, etc) of the model,
by minimizing the potential.
The FT (for example with respect to the $\mu$ term) is defined as
\cQ{$${\rm FT}(\mu)\equiv \frac{\mu^2}{M_Z^2}\frac{\partial M_Z^2}{\partial \mu^2}$$}
{${\rm FT}(\mu)\equiv (\mu^2/M_Z^2)(\partial M_Z^2/\partial \mu^2)$}
and tells how sensitive is the $Z$-mass with respect to variation of the parameters.

The result of a FT analysis in `minimal' supergravity (`mSuGra', sometimes also called
`constrained' suGra) MSSM
is the following\cite{FTLEP}:
the most recent bounds require a FT greater than about 6.
Is this an answer to our questions?
We believe not, due to the following reasons
\cQ{\begin{enumerate}}{}
\cQ{\item}{(1.)} Replacing $M_Z^2\to M_Z$ in the definition of the FT would reduce the minimal FT from 6 to 3.
\cQ{\item}{(2.)} Replacing $\mu^2\to \mu$ would increase the minimal FT from 6 to 12.
\cQ{\item}{(3.)} Should we worry if FT $> 5$, or 10, or 20, or\ldots?
\cQ{\item}{(4.)} The FT can be large in some cases  where nothing of unnatural happens\cite{FTcritica}.
\cQ{\end{enumerate}}{}
So,  the normalization of the FT is arbitrary and
setting an upper bound on the FT is a subjective choice.
Summarizing, a FT analysis only gives the following not very interesting statement:
\cQ{$$\hbox{
If we change the parameters of the mSuGra model by $1\%$,
the $Z$ mass changes by more than 3\%.}$$}
{``If we change the parameters of the mSuGra model by 1\% the $Z$ mass changes by more than 3\%''}
that does not seem to indicate a problem for supersymmetry.

\section{Naturalness and the allowed $\%$ of parameter space}
Since the FT does not give a useful answer, we approach the naturalness problem in a more direct way\cite{nat}t.
We will illustrate the question with various plots, all generated by the following procedure:
we consider the model that we want to study and

\begin{table}[t]
\caption{How naturally 3 different models generate the EW scale,
using 2 different ways of quantifying naturalness.\label{tab:exp}}
\vspace{0.4cm}
\begin{center}
\begin{tabular}{c|ccc}
&\Blue ElectroWeak scale  & Fine-tuning & acceptable \% of \Black\\ 
&\Blue $M_Z^2$ generated as & parameter & parameter space \Black\\ \hline
\Red Standard Model\Black & $\sum \frac{\alpha}{4\pi}(\Ord(\Lambda^2)+M^2)$&$\sim 10^{30}$&$\sim 10^{-30}$\\
\Red Techni-colour\Black   & $\Lambda^2 e^{-4\pi b/\alpha}$ & $\sim 50$~ & $\sim 1$\\
\Red MSSM (mSuGra)\Black  & $9 M_{1/2}^2-2\mu^2$ & $>6 $& $\sim 5\%$\\ 
\end{tabular}
\end{center}
\end{table}

\begin{enumerate}
\item We choose random values of $\Ord(1)$ for all the dimensionless ratios of soft parameters
(like $\mu/M_{1/2}$);
\item We do not impose how heavy are the sparticles; the overall mass scale is instead fixed
by requiring that the the minimum of the one-loop effective potential $V(h)$
be at its physical value $M_Z=g_2\langle h\rangle=91.18\GeV$;

\item Now everything is fixed and everything can be computed.
Since we want to study naturalness we check if the obtained spectrum of sparticles is sufficiently
heavy for satisfying all accelerator bounds;

\item We repeat the previous steps many times and we compute
$\%\equiv \frac{\hbox{number of acceptable spectra}}{\hbox{total number of spectra}}$.

\end{enumerate}
This Monte Carlo-like procedure computes 
how frequently numerical accidents can make the $Z$ boson
sufficiently lighter than the unobserved supersymmetric particles.
The main result is shown in table~1, where we compare
how natural are different models according to the FT and to our $\%$.
Both metods agree that the SM is perfectly unnatural in presence of high mass scales $M,\Lambda\circa{>}10^{16}\GeV$.
Fine-tuning says that techni-colour models generate the $Z$ scale in an unnatural way\cite{FTcritica},
while this is not true and our $\%$ is of order one
(concrete models would have various problems).
{\em The FT does not give a clear answer about the naturalness of the mSuGra MSSM ---
our procedure says that roughly $95\%$ of its parameter space
(with properly broken EW symmetry) is now excluded.}

Before analizing more models in more detail,
we comment about the solidity of the procedure that we follow.
It can be viewed as an efficient way of making plots with density of points {\em proportional} to 1/FT:
thus our $\%$ does not depend on the largely subjective normalization of the FT.
The $\%$ is normalized to be 100\% in absence of experimental bounds.
So a small $\%$ has a clear meaning.
Only {\em one aspect of the definition of the $\%$ is subjective\/}: in step~1 we must specify what we mean
with `order-one'.
In more general terms, we sample the parameter space according to some arbitrary
distribution of probability $p(\wp)$.
The results (like the allowed percentage of parameter space) depend on the choice of $p(\wp)$.
Actually this is a standard situation in inferential statistics:
for example an experimentalist that wants to report a lower bound on a neutrino mass $m_\nu$
needs to assume an arbitrary distribution of probability $p(m_\nu)$
(this assumption is explicit in the Bayesian approach\cite{Bayes}, where $p$ is called `prior distribution';
the situation is not better in the frequentistc approach, where different `confidence
intervals' are obtained using different techniques).


In conclusion the procedure that we will use
\cQ{\begin{itemize}}{}
\cQ{\item}{~~$\bullet$~~} is more intuitive (and simpler to compute) than FT;
\cQ{\item}{~~$\bullet$~~} does not tell that natural situations are unnatural;
\cQ{\item}{~~$\bullet$~~} does not tell that unnatural situations are natural
(differently from our $\%$, the FT only looks at how the $Z$ mass is obtained:
but other unnatural situations can be present, as we will exemplificate);
\cQ{\item}{~~$\bullet$~~} is not more subjective than --- say --- a 90\% C.L. bound on a neutrino mass.
\cQ{\end{itemize}}{

}
In short, our procedure is a sort of Bayesian analysis with an unconventional use of the
prior distribution: the only subjective aspect of our analysis is the
usual one, that no one knows how to to avoid.
So the real question is if this procedure is safe enough
(i.e.\ is there a decent choice of $p(\wp)$ that transforms the 5\% in table~1 into 50\%?):
we will discuss this question in the next section when we analyze concrete models.

\section{The supersymmetric naturalness problem}\label{masses}
In this section we discuss how serious is the supersymmetric naturalness problem
in the various different motivated scenarios of SUSY breaking.
We will not consider models with extra fields at low energy beyond the minimal ones present in the MSSM
and we will assume that $R$ parity is conserved.
To be conservative {\em we will consider as excluded only those spectra that violate the experimental
bounds coming from direct searches\/} at accelerators.

\subsection{A simple example}\label{example}
\begin{figure}[t]
\caption[SP]{
Fig.~1a: the naturalness problem in a simple supersymmetric model:
values of $\mu/M_2$ marked in dark gray are unphysical,
while \Orange light gray\Black{} regions have too light sparticles.
Only the small white vertical band is now experimentally acceptable.
Fig.~1b: Scatter plot of chargino mass vs higgs mass 
done with sampling density proportional to the naturalness probability
in minimal supergravity.
Only the \Green light gray\Black{} sampling points are allowed.
The shading covers excluded regions.
Fig.~1c: naturalness distribution of chargino mass:
allowed spectra give the $\sim 5\%$ tail in \GreenDark dark gray\Black{}.
\label{fig:example}}
\begin{center}
\begin{picture}(17,5)
\putps(-0.5,-1)(-0.5,-1){figMoriond}{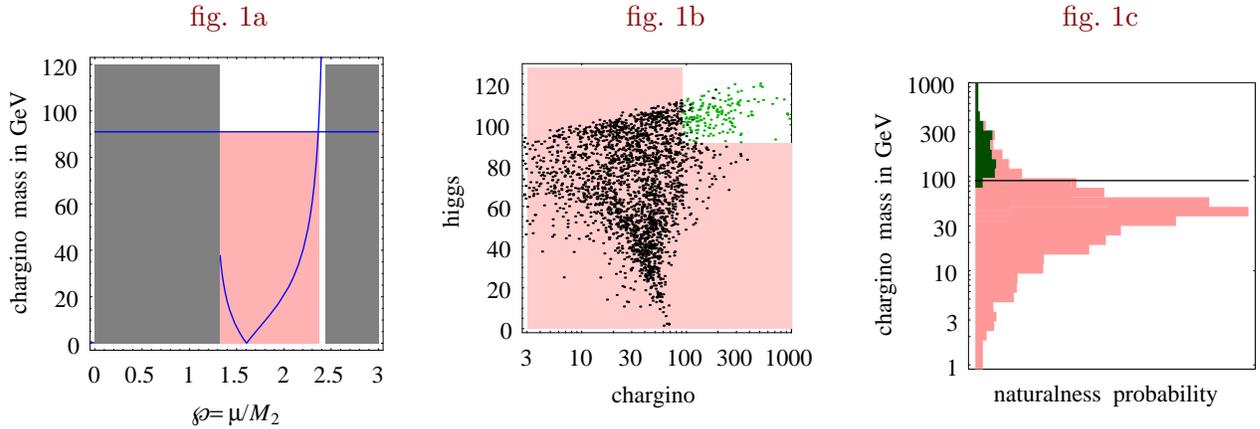}\Red
\put(2.5,4.5){fig.~1a}
\put(8.3,4.5){fig.~1b}
\put(14.1,4.5){fig.~1c}\Black
\end{picture}
\cQ{\vspace{5mm}}{}
\end{center}\end{figure}

We now try to illustrate the previous discussions with a simple and characteristic example.
We consider the `most minimal' gauge mediation scenario with very heavy messengers
(one $5\oplus\bar{5}$ multiplet of the unified gauge group SU(5) with mass $\MM=10^{15}\GeV$).
We also assume that the unknown mechanism that generates the $\mu$-term
does not give additional contributions to the other parameters of the higgs potential.
Since the $B$ term vanishes at the messenger scale,
we obtain moderately large values of $\tan\beta\sim20$.
This model is a good example because it is not completely irrelevant and
\begin{itemize}

\item its spectrum is not very different from
a typical supergravity spectrum and is the most natural possibility within
`conventional' models.
It has the lowest FT (we fix the unprecisely measured parameter
$\lambda_t(\MGUT)$ at a small value (0.5)
that alleviates the naturalness problem).

\item it is simple because it has only two free parameters:
the overall scale of gauge mediated soft terms and the $\mu$ term.
The condition of correct electroweak breaking fixes the overall mass scale,
and only one dimensionless parameter remains free.
We choose it to be $\wp\equiv \mu/M_2$ (renormalized at low energy),
where $M_2$ is the mass term of the $\SU(2)_L$ gaugino.

\end{itemize}
In figure~1a we plot, as a function of $\wp$,
the lightest chargino mass in GeV.
The dark gray regions where $\mu$ is too small or too large are always excluded from our analysis
because the electroweak gauge symmetry cannot be properly broken.
Values of $\wp$ shaded in \Orange light gray\Black{}
are excluded because some supersymmetric particle is too light
(in this example the chargino gives the strongest bound).
The chargino is heavier than its LEP2 bound
only in small ranges of the parameter space at $\mu/M_2\approx  2.3$ (left unshaded in fig.~\ref{fig:example})
close to the region where EW symmetry breaking is not possible
because $\mu$ is too large.
{\em The fact that the allowed regions are very small and atypical
is a naturalness problem for supersymmetry}.
About $90\%$ of the acceptable values of $\wp$ are now excluded
by the chargino mass bound.
This percentage would not be significantly different if we had used $\mu^2$ in place of $\mu$ as a parameter;
it could be significantly higher only choosing a rather ad-hoc parametrization.
Indirect bounds (that we neglect) are not completely negligible:
about half of the allowed parameter space of this example is excluded
because the $b\to s\gamma$ decay is too fast.

\subsection{Minimal supergravity}
``Minimal supergravity'' assumes common values $m_0$, $M_{1/2}$ and $A_0$
at the unification scale $\MGUT\approx2\cdot10^{16}\GeV$
for the sfermion masses,
the three gaugino masses, and the $A$-terms.
The parameters $\mu_0$ and $B_0$ are free.
\cQ{Since we adopt standard notations, we do not list the exact definitions of
the various supersymmetric parameters.}{}
As explained in the previous section we randomly fix the dimensionless ratios
of the soft terms
and compute the overall supersymmetric
mass scale ``$m_{\rm SUSY}$'' from the minimization condition of the MSSM potential.
More precisely we scan the parameters
within the following ranges
$$m_0,|\mu_0|,|M_{1/2}|,|B_0|,|A_0|=(0\div 1)m_{\rm SUSY},\qquad
(a\div b)\equiv \hbox{a random number between $a$ and $b$}.$$ 
The alternative scanning procedure
\cQ{$$m_0=(\frac{1}{9}\div 3) m_{\rm SUSY},\qquad
|\mu_0|,|M_{1/2}|=(\frac{1}{3}\div 3) m_{\rm SUSY},\qquad
A_0,B_0=(-3\div 3) m_0$$}
{$m_0=(\frac{1}{9}\div 3) m_{\rm SUSY}$,
$|\mu_0|,|M_{1/2}|=(\frac{1}{3}\div 3) m_{\rm SUSY}$,
$A_0,B_0=(-3\div 3) m_0$}
(with the samplings of $m_0, M_{1/2}$ and $\mu_0$ done with flat density in logarithmic scale)
would give the same final results: {\em 
\cQ{in the MSSM with soft terms mediated by `minimal supergravity'}{}
the present experimental bounds exclude $95\%$ of our parameter space.\/}

As in the previous example, reasonable different definitions of `order one'
(i.e.~of the prior Bayes distribution)
do not alleviate the naturalness problem:
we could make the naturalness problem apparently more dramatic 
by restricting the dimensionless ratios to a narrow region 
that does not include some significant part of the experimentally allowed region, 
or by extending the range to include larger values that produce a larger spread 
in the spectrum so that it is more difficult to satisfy all the experimental bounds. 
Allowing one or few soft terms to be much smaller or much larger than the others does not help.
Complex soft terms only give problems with electric dipoles.

\smallskip

\cQ{We now exhibit the results of this analysis in a series of figures.}{}
In fig.~1b we show a scatter plot
with sampling density proportional to the naturalness probability.
The sampling points corresponding to excluded (still allowed) spectra are drawn in black,
(in \Green light gray\Black{}).
Fig.~1b shows the correlation between the masses of the lightest chargino
and of the lightest higgs, $m_\chi$ and $m_h$.
We see that the most natural value for the chargino mass is around $30\GeV$
(rather than $300\GeV$):
however numerical accidents can sometimes make the chargino so light that it can be produced
at $B$-factories, or make it so heavy that it cannot be produced at LEP2.
Fig.~1b also shows that the experimental bounds on $m_\chi$ and $m_h$
are the most important ones in minimal supergravity
(and that the bound on $m_\chi$ is more important than the one on $m_h$).
The LEP2 bounds together with the assumption of gaugino mass universality at the unification scale 
give stronger constraints on the mass of coloured sparticles than the direct bounds from Tevatron experiments.


In fig.~1c we show the same kind of results using a different format:
we plot the density of the points in fig.~1b as function of $m_\chi$.
We obtain a bell centered around $30\GeV$.
Allowed spectra are confined in fig.~1c to
the small ($\sim 5\%$) upper tail where the chargino is anomaously heavy.

\subsection{`Unified' supergravity}
If we only impose unification relations between soft terms,
the naturalness problem remains unchanged:
an unnatural cancellation remains necessary even if there are more parameters.
From the point of view of phenomenology, maybe the most interesting 
new possibility is that a mainly right handed stop
can `accidentally' become significantly lighter than the other squarks.
A very light stop can mediate large loop effects (in $b\to s\gamma$, $\Delta m_B$, $\eps_K$, \ldots)
and allows baryogenesis at the weak scale\cite{EWbaryogenesis}.
However this possibility is severely limited by naturalness considerations\cite{lightStop,nat}:
it requires not only that one numerical accident makes the $Z$ boson
sufficiently lighter than the unobserved chargino,
but also that a second independent numerical accident gives a stop lighter than the other squarks.
This unnatural combination occurs very rarely ($p\sim10^{-3}$) in our analysis.

\subsection{Gauge mediation}
For our purposes `gauge mediation' models\cite{GaugeSoft} can be characterized
saying that they predict the 
soft terms $M_i$, $m_R$, $A$ of sparticles
in terms of their gauge charges $c_R^i$ in the following way:
$$
M_i(\MM)=\frac{\alpha_i(\MM)}{4\pi} M_0,\qquad
m_R^2(\MM) =  \frac{c^i_R }{\sqrt{n}}M_i^2(\MM) ,\qquad
A(\MM)=0
$$
($R$ runs over the sfermions and $i$ over the gauginos).
These predictions depend on three unknown parameters:
the overall scale of soft terms $M_0$,
the messenger mass $\MM$ and the `number of messengers' $n$.
A computation shows that:
\cQ{\begin{itemize}}{}
\cQ{\item}{~~$\bullet$~~} If the messengers are heavy, $\MM\circa{>}10^{12}\GeV$
the gauge mediated sparticle spectrum, and the naturalness problem,
are not much different than in supergravity.
\cQ{\item}{~~$\bullet$~~} If $\MM\sim 10^{(6\div 10)}\GeV$ gauge mediation models predict a right handed
selectron mass significantly smaller than the higgs mass term which sets
the scale of electroweak symmetry breaking.
This makes the naturalness problem more acute\cite{GMFT}.
\cQ{\item}{~~$\bullet$~~} If the messengers are very light ($\MM\circa{<}100\TeV$)
and the lightest neutralino $N$ decays in a detectable way ($N\to \gamma\,\hbox{gravitino}$)
within the detector,
experimentalists can put a {\em very\/} strong bound on its mass
that makes the model very unnatural\cite{nat}.
\cQ{\end{itemize}}{}
\cQ{Extremely light messengers ($\MM\approx 10\TeV$)
could give rise to a more natural sparticle spectrum, but unknown NLO corrections,
that depend on unknown couplings between messengers, become relevant in this limiting case.}{}

\section{The credibility of supersymmetric models after LEP2}
If we employ the standard notation
\cQ{$$}{$}
p(A|B)\equiv \hbox{(probability of $A$ assuming that $B$ is true)}
\cQ{$$}{$}
we can compactely rewrite our main result as
\cQ{\begin{equation}}{$\Blue}
p(\LEP|\SUSY)=5\%\label{eq:5}
\cQ{\end{equation}}{$\Black}
where \LEP{} and \SUSY{} are abbreviations for
\begin{eqnarray*}
\LEP &=&\hbox{``no supersymmetric signal found at LEP and Tevatron experiments''}\\
\SUSY &=&\hbox{``MSSM with minimal supergravity, with comparable soft terms''}
\end{eqnarray*}
Our result means that if \SUSY{} is true, it is strange that
sparticles have not been discovered at LEP\footnote{We repeat that {\bf \SUSY{} does not mean supersymmetry\/} ---
it is just a particular representative model often used for discussing capabilities
of accelerators and for other purporposes (we are also assuming that soft terms are comparable,
and that there is no special correlation between them).
However a similar naturalness problem is present in all `conventional' supersymmetric models.}.
We do {\bf not} claim that $p(\SUSY|\LEP)=5\%$, i.e.\ that \LEP{} exclude \SUSY{} at $95\%$ C.L.
Nothing says that the inverse probability $p(\SUSY|\LEP)$
is equal to the direct one $p(\LEP|\SUSY)$.
However we can say something on the more interesting inverse probability because
the Bayes theorem\cite{Bayes}
says how they are related:
$$\frac{p(\SUSY|\LEP)}{p(\SUSY)}=\frac{5\%}{1-95\%p(\SUSY)}$$
In order to apply the theorem we have introduced an additional quantity
$p(\SUSY)=$ degree of belief that \SUSY{} (rather than some other
unspecified mechanism) is the solution to the SM naturalness problem
before knowing the experimental results.

The theorem says how much one must reduce its confidence in \SUSY{} after knowing the LEP2 results.
Of course it gives obvious results.
For example who very strongly believes in $\SUSY$ would insert $p(\SUSY)=1$
and get $p(\SUSY|\LEP)=1$: experiments are explained by saying that we live in the
small allowed corner of the parameter space of \SUSY{} with heavy sparticles.
Who believes strongly in supersymmetry could instead suspect that the true model has not yet been found.

\section{Conclusions}\label{fine}
In conclusion, the negative results of the recent searches for supersymmetric particles
pose a naturalness problem to all `conventional' supersymmetric models.
What could be the reason of these negative experimental results?
\cQ{\begin{itemize}}{}
\cQ{\item}{~~$\bullet$~~}  Maybe SUSY has escaped detection beacuse a numerical accident
has made the sparticles too heavy for LEP2.
This happens with $\sim5\%$ probability.
Once one accepts the presence of a numerical accident,
it is no longer extremely unlikely that the coloured sparticles have mass of few TeV (beyond the LHC reach) due to an
accidental cancellation stronger than the `minimal one'~\cite{nat};
\cQ{\item}{~~$\bullet$~~} Maybe SUSY has not been discovered because it does not exists at the weak scale;
\cQ{\item}{~~$\bullet$~~} Maybe SUSY exists at the weak scale, there is no unluck accident, but
some ingredient is missing in the `conventional' models.
\cQ{\end{itemize}}{

}
Some recent papers have studied few different possible ways of alleviating the naturalness problem:
\cQ{\begin{enumerate}}{}
\cQ{\item}{(1.)} A correlation between the $\mu$ term and the gaugino mass\cite{corr}.
However this correlation reduces the FT with respect to $\mu$, but not the FT
with respect to $\lambda_{\rm top}$ or to $\alpha_{\rm strong}$.
The problem is that the required value of $\mu/M_{1/2}$ depends on the values of
$\lambda_{\rm top},\alpha_{\rm strong},\tan\beta\ldots$
(i.e.\ even if the high-energy theory predicts a $Z$ boson ligher than sparticles,
the prediction is destabilized by large RGE corrections).
\cQ{\item}{(2.)}
Models with SUSY breaking mediated at lower energy could predict a light $Z$ boson,
not destabilized by quantum corrections.
However concrete models of this kind (gauge mediation\cite{GaugeSoft}
and higher dimensional Scherk-Schwarz\cite{SS}\footnote{The predictions
of these models depend on the form of the ultraviolet cutoff;
even for the cutoff used in\cite{SS} the dependence of the higgs potential
on the arbitrary matching scale
(beteween the full theory and the low energy 4-dimensional theory)
is so strong to require a NLO computation, which
result would depend on the unknown details of the high energy theory.} models)
turn out to be not more natural than supergravity.
\cQ{\item}{(3.)}
A sparticle spectrum more degenerate than the `conventional' one.
Since the Tevatron direct bound on the gluino mass is weaker than the indirect bound
obtained from LEP2 assuming gaugino mass universality,
it is possible to reduce the mass of coloured particles
(and consequently their large RGE corrections to the $Z$ mass, that sharpen the naturalness problem)
if gaugino mass universality is strongly broken\cite{FTred,bau,nat}.
The phenomenology is more interesting than in `conventional' models:
the gluino is lighter (could be discovered at Tevatron) and
the minimal cross section for detecting a neutralino dark matter is $\sim 50$ times higher.
However even this case is severely constrained by experimental bounds:
often at least one sparticle happens to be lighter than the $Z$ boson
and contradics one of the many experimental bounds.
\cQ{\end{enumerate}}{

}
In conclusion  it is certainly possible to debate if there is a supersymmetric naturalness problem,
sometimes with considerations that also make the SM natural,
thus killing the motivation for supersymmetry.
Once that the problem is accepted it is however a solid problem:
as usual it cannot be eluded with mild assumptions about the high energy theory.

\Black

\end{document}